\documentclass[12pt]{article}
\usepackage{graphicx}
\usepackage{amssymb}
\newcommand{\beq}{\begin{equation}}
\newcommand{\eeq}{\end{equation}}
\newcommand{\beqn}{\begin{eqnarray}}
\newcommand{\eeqn}{\end{eqnarray}}

\begin{document}

\begin{center}
{\Large \bf  Comment on "Comments on recent work\\
\vspace{3mm} on dark-matter capture in the Solar System"}
\end{center}

\begin{center}
I.B.~Khriplovich\footnote{khriplovich@inp.nsk.su}\\
Budker Institute of Nuclear Physics\\
630090 Novosibirsk, Russia,\\
\end{center}

\begin{abstract}
The criticism contained in the recent preprint arxiv:1004.5258 is based
essentially on misquoting the articles criticized therein. As to the
conclusion advocated in that preprint, according to which the density
of dark matter bound to the Solar System is small as compared to the
dark-matter density in the Galactic halo, it is not clear whether this
claim is correct.
\end{abstract}

In the recent preprint \cite{ep}, Edsj\"{o} and Peter claim
that in the papers by Khriplovich and Shepelyansky \cite{ks} and
by Khriplovich \cite{kh} "the authors find huge enhancements of the
density of dark matter bound to the Solar System compared to the
local dark-matter density in the Galactic halo".

Let us compare this assertion with the true conclusions
made by Khriplovich and Shepelyansky in \cite{ks} and by Khriplovich
in \cite{kh}.

As regards the article \cite{ks}, our estimate for the density of dark matter
captured by the Solar System, presented therein, is very high
indeed, on the level of $10^{-21}$ g/cm$^3$. However, we state
explicitly in the article the origin of this estimate, our own attitude
to it, and our aim of presenting thus obtained numbers: "... we
can make an assumption resulting in the most optimistic prediction
for the "partial" {\em(i.e. captured by a given planet)} dark matter
densities $\Delta \rho_p$. We assume that each of the total masses
$\Delta m_p$ of the captured dark matter occupies the volume
$(4\pi/3)r^3_p$, where $r_p$ is the orbit radius of the
corresponding planet. We do not claim that this assumption is
correct, but believe, however, that the comparison of its (almost
certainly, overoptimistic) results with the observational limits
{\em(on the dark matter density)} will be instructive."

As to \cite{kh}, my estimate made therein for the density of dark
matter, captured by the Solar System, in the vicinity of the
Earth:
\[
\rho\sim 5 \cdot 10^{-25}\;\,{\rm g/cm^3}\,,
\]
practically coincides with the common value
\[
\rho_g \simeq 4 \cdot 10^{-25}\;\rm{g/cm}^3
\]
for the galactic dark-matter density. Thus, as to \cite{kh}, its
misquotation by Edsj\"{o} and Peter \cite{ep} is obvious.

In conclusion, I address the issue crucial for the estimate of the
density of dark matter captured by the Solar System. This is the
problem of the characteristic time of the inverse process, i.e.
that of the ejection of the captured dark matter from the Solar
System. The arguments in favor of the short life-time of the
captured dark matter in the Solar System are advocated by Edsj\"{o}
and Peter \cite{ep}. There are however the
arguments indicating that this life-time is comparable to, or even
larger than, the life-time of the Solar System presented by
Khriplovich and Shepelyansky \cite{ks}. Thus, contrary to the
assertion by Edsj\"{o} and Peter \cite{ep}, the situation with the
true value for typical life-time of the captured dark matter in the Solar
System is far from being clear.
\vspace{3mm}

{\bf Acknowledgement.} The work was supported by the Russian
Foundation for Basic Research through Grant No. 08-02-00960-a.


\begin{thebibliography}{99}

\bibitem{ep}
J. Edsj\"{o} and A.H.G. Peter, arxiv:1004.5258 (astro-ph.EP)
(2010).

\bibitem{ks}
I.B. Khriplovich and D.L. Shepelyansky, Int. J. Mod. Phys. D
\textbf{18}, 1903 (2009); arxiv:0906.2480 (astro-ph.SR) (2009).

\bibitem{kh}
I.B. Khriplovich, arxiv:1004.3171 (astro-ph.EP) (2010).

\end{thebibliography}
\end{document}